\documentstyle[12pt]{report}
\begin{document}
\large
\begin{center}{\bf THE CERES ASTRONOMICAL DATABASE
       }\end{center}  
\normalsize
\begin{center}{E. Xanthopoulos, N. J. Jackson \\ {\it Jodrell Bank Observatory, 
United Kingdom}}\end{center} 
\begin{center}{I. Snellen \\ {\it IoA, University of Cambridge, United Kingdom}}\end{center}
\begin{center}{J. Dennett-Thorpe \\ {\it Kapteyn Institute, Groningen, The Netherlands}}\end{center}
\begin{center}{K.-H. Mack \\ {\it Istituto di Radioastronomia del CNR, Bologna, Italy}}\end{center}  

\vspace{10mm}
\noindent
{\bf ABSTRACT}
\vspace{2mm}

The CERES Astronomical Archive was created initially in order to make the large amount of 
data that were collected 
from the two surveys, the Jodrell-VLA Astrometric Survey (JVAS) and 
the Cosmic Lens All-Sky Survey (CLASS), easily accessible to the partner members of the 
Consortium for European Research on Extragalactic Surveys (CERES) through the web. 
Here we describe the web based archive of the 15,000 flat spectrum radio sources.
There is a wealth of information that one can access including downloading the raw 
and processed data for each of the sources. We also describe the search engines that 
were developed so that a) one can search through the archive for an object or a sample
of objects by setting specific criteria and b) select new samples. Since new data are 
continually gathered on the same sources from follow up programs, an automatic 
update engine was created so that the new information and data can be added easily in the archive. 

\vspace{5mm}

\noindent
{\bf INTRODUCTION}

\vspace{2mm}

The Consortium for European Research on Extragalactic Surveys (CERES) is a TMR research network
that was created in order to work on two major surveys the Jodrell-VLA Astrometric Survey (JVAS; 
Patnaik et al. 1992; Patnaik 1993; Browne et al. 1998, Wilkinson et al. 1998)
and the Cosmic Lens All-Sky Survey (CLASS; Browne et al. 1998) which together contain a total of 
$\approx$ 15,000
flat spectrum radio sources. Initially the main research objective was to find new gravitational
lens systems (17 lenses have been found up to now),
and to use the unbiased samples of gravitational lenses 
for constraints on cosmological parameters. CERES was also set up to study high and low
redshift AGN. 

The survey consisted of observations of all the flat spectrum radio sources in the Northern sky 
(declination between 0$^\circ$ and 75$^\circ$, galactic latitude $\ge$ 10$^\circ$, 
spectral index $\ge -0.5$ 
based on the flux densities in the GB6 (Gregory et al. 1996) and the NVSS (Cotton et al. 1996) 
surveys) that had a GB6~5 GHz flux 
density $\ge$ 30 mJy. The survey was done in different epochs (called JVAS 1, 2 and 3 and 
CLASS 1, 2, 3 and 4) but a recalibration of all the data was performed in 1999 in order 
to achieve a uniform data reduction and sample. 
Following the above criteria 10,499 sources compiled the so-called ``statistically complete
sample" which forms the basis of the archive, while the rest of the observations that do not
follow the above criteria 
comprise the so-called ``supplementary sample".     
The sources in the ``statistically complete sample" were named following  
the naming convention of the GB6 catalogue (J2000 coordinates).
For the supplementary sample the sources
are known by names of their original samples which are based on two different naming systems 
depending on when they were selected.  

In order to facilitate the use of the initial
and follow up data by the CERES partner members and 
by the scientific world when the archive will become publicly available, 
an archive/database was created that holds
all the data and background information. 
Publicly available information can be found at the URL address: \\
http://www.jb.man.ac.uk/$\sim$ceres1.

\vspace{5mm}
\noindent
{\bf THE DATA ARCHIVE} 

\vspace{2mm}

As soon as one enters the data archive one has the option to link to: 
\begin{itemize}
\item[\bf a.] the statistically complete sample
\item[\bf b.] the supplementary sample
\item[\bf c.] the statistically complete sample using the initial survey names
\end{itemize}
{\bf A. The statistically complete sample}

\vspace{2mm}

10,499 webpages have been created each containing all the available information for
each source in the ``statistically complete sample".
By clicking on the ``statistically complete sample" one has the option to 
{\bf a)} download the whole sample webpages {\bf b)} download the  webpages 
from sources ranging over a 1 hour 
RA interval or
{\bf c)} download a more simple webpage format of the whole sample.

By clicking on a specific GB6 name/source one can then access a webpage of the format 
shown in Figure 1 (the figure has been created from a combination of the webpages of 
two sources so that it shows all the possible
information that can be available from the archive).

Information and data that are available for each source are as follows:
\begin{itemize}
\item{\bf General Information}: This includes the official GB6 J2000 and B1950 name for 
each source obtained from the GB6 position in the manner described in Gregory et al. 1996, 
as well as J2000 and B1950 coordinates from the initial pointing (NVSS, WENSS,
TEXAS) that were available at the time of observations, the GB6 5 GHz coordinates and 
finally the CLASS 8.4 GHz coordinates, with an accuracy of 200 mas,
that came out from the recalibration of all the CERES data.
\item{\bf Flux densities}: For each source one can find the WENSS (0.325 GHz), TEXAS (0.365 GHz),
NVSS (1.4 GHz) and GB6 (4.85 GHz) flux densities in mJy, if available. 
One has also the option to download a ``flux map".
By providing the coordinates of your object, a cross correlation with all the available 
catalogues is performed and the ``flux map" is returned. This gives all the flux densities 
known for that
source and their positions. This map enables the validity of the identifications of sources
in different catalogues to be checked at a glance.   
\item{\bf Radio map information}: This entry supplies the information that we have from 
the initial radio data of the survey and namely a) the 8.4 GHz flux density derived from the VLA 
data b) the old observation name by which the source was known at the time of the observations
c) the date of the observations d) the number of visibilities e) the CLASS epoch (when the 
source was observed, JVAS 1 2 or 3, CLASS 1, 2, 3 or 4) f) the CLASS tape number (this 
refers to the archive tapes kept at Jodrell Bank).
\item{\bf Flags}: 
Other information relevant to the radio properties of the target can be obtained with flags.
This includes:
a) The 1.4 GHz NVSS map of the source consisting of a 4 arcmin diameter circle around the GB6 position 
which can be downloaded by clicking on the button, 
b) one can also check whether from the data reduction 
it was found that the source has multiple components or c) the number of fields processed in 
AIPS (these reduced maps can also be downloaded by clicking on the appropriate number) 
and d) a quality control factor for each map that is defined as the ratio of 
the peak flux density before the selfcalibration 
divided by the final peak flux density (this is the reliability number). Empirically experience 
shows that the results on sources with reliability numbers $>$2 should be regarded with caution.   

\item{\bf Optical information}: There are two main sources from which we get all the 
optical information: {\bf a)} our targeted optical follow up observations of the sources that give 
1. the morphological type, 2. the redshift, 3. the redshift error, 4. the date of 
observation, 5. the names of the observers and reducers, 6. the lines that were used in the spectra
for determining the redshift, and 7. any specific notes for the object from 
the optical observations; {\bf b)} from the APM archive we get the following information for 
each source: 1. the optical offset in the optical position given our radio position, 2. the 
APM R ID (galaxy, stellar etc.), 3. the APM PSF in the Red, 4. the R magnitude, 5. the 
APM B ID, 6. the APM PSF in the Blue, 7. the B magnitude, 8. the colour (B-R), 9. the RA and 10.
the DEC J2000 optical coordinates.
\item{\bf The data}: Actual data that now can be downloaded: 1. VLA 8.4 GHz 
A-configuration raw data (in FITS format), 2. VLA 8.4 GHz maps (in gif format), 3. VLA 8.4 GHz maps
(postscript compressed files), 4. APM map (4$\times$4 arcmin in gif format), 5. APM list 
(text file that identifies the sources seen in the APM map. There is also a direct link 
to the DSS (Digital Sky Survey) with automatic input of the coordinates. One has only  
to select the type of format of the data and the size of the map and click to download.
There are more entries for future MERLIN, HST, optical spectra, X-ray and other wavelength data 
that will be available to download. 
\end{itemize}
\noindent
{\bf B. The supplementary sample}

\vspace{2mm}

A separate webpage archive is created for all the observations that do not fall in the 
``statistically complete sample" as defined using the criteria above. The webpages have 
the same format and type of information described above, where available. 

\vspace{5mm}
\noindent
{\bf THE SEARCH ENGINES}

\vspace{2mm}

In order to search for what is in the database for specific objects and also to select
new samples two search engines have been developed:

\vspace{2mm}
\noindent
{\bf 1. Selection Engine:}
The web-interface allows the user to select sources on their
radio position, flux densities in different surveys, spectral indices,
redshift and optical parameters. It selects sources from the masterlist-file, a text file with a  
collection of all the information that we have for all the sources in our archive. 
It returns those sources from the masterlist with the 
specified criteria, giving all the parameters available
(class name, GB6 positions, WENSS, NVSS, GB6, CLASS flux densities, APM R and B magnitudes,
redshift, morphological type, CLASS and NVSS positions as well as links 
to available optical spectra or images and VLA maps). 
In addition it provides links to the individual source-webpages,
to NVSS images, DSS images, to other databases like NED and APM,
and also the option of a text-file  output version.

\vspace{2mm}
\noindent
{\bf 2. Matching-list Engine:}
This feature allows the user to see whether a list of objects
is in the database, by entering a list of GB6 names, J2000 positions or both.
This list is checked against the masterlist (using a user defined
search-radius). The outcome is similar to the above.

\vspace{5mm}
\noindent
{\bf AUTOMATIC UPDATING}

\vspace{2mm}

Since we plan to add new data from radio and optical observations, in order to automate everything,
{\bf an ``update" program} has been developed. 
An authorised user can use this programme to add new data to the database. The new data may be
one of several classes (e.g. redshift, new optical or radio images, spectra, or comments) which the
user is required to declare, and the input formatted accordingly. The program then runs a series of checks 
and informs the user of any errors, before updating the webpages and associated lists.
Because of security problems with web-based programmes, we decided not to implement a web version of this
programme, as originally planned.

\vspace{5mm}
\noindent
{\bf References}

\vspace{5mm}
\noindent
Browne, I. W. A., Jackson, N. J., Augusto, P., Henstock, D. R., Marlow, D. R., Nair, S., Wilkinson, 
P. N., 1998. The JVAS/CLASS gravitational lens surveys, in Bremer, M. N., Jackson, N., 
P\'{e}rez-Fournon, I., eds, Observational Cosmology
with the new radio surveys, Astrophysics and Space Science Library, Vol. 226. 
Dordrecht: Kluwer Academic Publishers, p. 323

\vspace{2mm}
\noindent
Browne, I. W. A., Wilkinson, P. N., Patnaik, A. R., Wrobel, J. B., 1998. Interferometer phase 
calibration sources. II - The region 0$^\circ$$\le$ $\delta _{B1950}$ $\le$$+$20$^\circ$ , MNRAS, 293, 257

\vspace{2mm}
\noindent
Cotton, W. D., Condon, J. J., Yin, Q. F., et al. , 1996. The NRAO VLA D-Array Sky Survey (NVSS), 
in proceedings of the 175th Symposium of the International Astronomical Union, eds. Ron D. Ekers, 
C. Fanti, and L. Padrielli, Kluwer Academic Publishers, p. 503

\vspace{2mm}
\noindent
Gregory, P. C., Scott, W. K., Douglas, K., Condon, J. J., 1996. The GB6 Catalog of Radio Sources,
ApJS, 103, 427

\vspace{2mm}
\noindent
Patnaik, A. R., Browne, I. W. A., Wilkinson, P. N., Wrobel, J. M., 1992, MNRAS, 254, 655

\vspace{2mm}
\noindent
Patnaik, A. R., 1993, Proceedings of the 31st Li\`{e}ge International Astrophysical Colloquium 
``Gravitational Lenses in the Universe", p. 311

\vspace{2mm}
\noindent
Wilkinson, P. N., Browne, I. W. A., Patnaik, A. R., Wrobel, J. M., Sorathia, B., 1998, MNRAS, 300, 
790
\end{document}